\documentstyle[11pt,paspconf,epsf]{article}
\font\sit   = cmti9                  
\def\parano    {\par \noindent}

\def\Mej{M_{\rm ej}}

\def\EDM{Ellison, Drury, \& Meyer }
\def\EMP{Ellison, M\"obius, \& Paschmann }
\def\HLR{Higdon, Lingenfelter, \& Ramaty }

\def\LRK{Lingenfelter, Ramaty, \& Kozlovsky }
\def\MDE{Meyer, Drury, \& Ellison }
\newcount\listno
\def\list{\global\advance \listno by 1 {(\the\listno) }}
\def\newlist{\listno=0}
\def\alf{Alfv\'en }
\def\al26{$^{26}\!$Al}

\def\Vg{v_{\rm G}}

\def\kmps{km s$^{-1}$}
\def\x#1{\times 10^{#1}}

\def\egc{e.g., }
\def\ie{i.e. }
\def\iec{i.e., }
\def\etal{et al.~}
\def\etalc{et al., }

\def\pcc{cm$^{-3}$}
\def\DiffL{l_D}
\def\nH{n_{\rm H}}
\def\tloss{t_{\rm loss}}
\def\tG{t_G}
\def\mp{m_p}

\def\Lgrain{\lambda_G}
\def\tloss{t_{\rm loss}}
\def\Vsk{V_{\rm sk}}
\def\Rsk{R_{\rm sk}}
\def\tSNR{t_{\rm snr}}
\def\Mswept{M_{\rm swept}}
\def\Msun{M_{\odot}}
\def\EnSN{E_{\rm sn}}
\def\solar{\odot}

\def\Ne22{$^{22}$Ne}
\def\Fe57{$^{57}$Fe}

{\catcode`\@=11
\gdef\SchlangeUnter#1#2{\lower2pt\vbox{\baselineskip 0pt\lineskip0pt
\ialign{$\m@th#1\hfil##\hfil$\crcr#2\crcr\sim\crcr}}}}
\def\gtrsim{\mathrel{\mathpalette\SchlangeUnter>}}

\newcount\eqnno
\def\eqnum{\global\advance \eqnno by 1 \eqno{(\the\eqnno)}}
\def\eqnumal{\global\advance \eqnno by 1 \eqno{(A\the\eqnno)}}
\def\label#1{\global\advance \eqnno by 1\xdef#1{\the\eqnno }
    \global\advance \eqnno by -1}
\def\edcomment#1{\iffalse\marginpar{\raggedright\sl#1\/}\else\relax\fi}

\reversemarginpar

\begin{document}

\title{\vskip-5mm The Origin of Present Day Cosmic Rays: \\
Fresh SN Ejecta or Interstellar Medium Material ? \\
II$\;\>$Physics of the SNR shock wave acceleration *}

\author{\vskip-6mm Donald C. Ellison}

\affil{\vskip-3mm Department of Physics, North Carolina State University, \\ Box 8202,
Raleigh, NC 27695, USA \\ don\_ellison@ncsu.edu}

\author{\vskip-1mm Jean-Paul Meyer}

\affil{\vskip-1.5mm Service d'Astrophysique, DAPNIA, CEA/Saclay, \\ 91191
Gif-sur-Yvette, France \\ meyer@hep.saclay.cea.fr}

\begin{abstract}
\newlist Assuming that refractory elements in cosmic rays
originate in dust grains, we examine the viability of cosmic ray
origin models wherein the bulk of {\it present day} cosmic rays are
accelerated out of fresh supernova ejecta material before it mixes
with the average interstellar medium (ISM).  We conclude that the
fresh ejecta scenarios that have been proposed thus far have serious
flaws, and are unable to account for known properties of present day
cosmic rays.  These flaws include:
\list the small fraction of ejecta to ISM mass processed by the
forward supernova remnant (SNR) shock;
\list the difficulty fresh ejecta grains have in reaching the forward
shock in isolated SNRs, and the small expected sputtering yield,
especially {\it ahead\/} of the shock, even
if grains do reach the forward shock;
\list the implausibility that fresh ejecta material can dominate
cosmic ray production in diffuse superbubbles; and
\list the lack of a connection in fresh ejecta models
between the production of cosmic ray refractory and volatile elements.
We conclude that the near linear increase in Be abundance with
metallicity observed in old, halo stars {\it cannot} imply that a
significant fraction of the cosmic rays {\it seen today} come from
fresh supernova ejecta.
This conclusion is supported by the analysis of the present day
cosmic ray composition, as shown in Meyer \& Ellison, this volume.
\end{abstract}

\vskip-12pt
\section{Introduction}

Observations strongly suggest that the source material for galactic
cosmic rays is a mixture of ISM gas and dust (\MDE 1997).  A
quantitative model that includes the shock acceleration of ISM and/or
circumstellar gas and dust by SNR blast waves (\EDM 1997) is able to
explain the abundance observations by making only standard assumptions
for nonlinear
%
\begin{minipage}{13.34cm}  
\vskip 0.25truecm 
\parano  -----------------------\vskip -0.2truecm
{\sit \parano {{\rm*} Workshop on "LiBeB, 
Cosmic Rays and Gamma-Ray Line Astronomy",
IAP, Paris,\break\vskip-5mm 9-11 December 1998, 
M.\ Cass\'e, K.\ Olive, R.\ Ramaty, \& E.\ Vangioni-Flam eds.,  
ASP Conf.\break\vskip-5mm Series 
(Astr.\ Soc.\ of the Pacific, 1999), in press.\par}}
\end{minipage}
\newpage

\noindent diffusive shock acceleration in SNRs, grain properties,
and grain interactions in the ISM.

Now, prompted by observations of the Be abundance
in old, low metallicity halo stars (see Vangioni-Flam \etal 1998 and
Fields \& Olive  1999 for
reviews and references),
\LRK (1998) have suggested that
both the {\it early galactic\/} Be evolution and the observed abundances of
{\it current\/} cosmic rays might be explained
if the refractory elements in cosmic rays come from the acceleration
of dust from {\it fresh supernova ejecta material}, rather than
from the surrounding average ISM (and possibly
the circumstellar) material.
The impetus for this comes from the observation that the Be
abundance in halo stars increases approximately linearly with
metallicity (\iec the Fe/H abundance ratio) from fairly early times
with [Fe/H] $\sim -3$ to
about [Fe/H] $\sim -0.5$.\footnote{[Fe/H] is the logarithmic
abundance ratio by number relative to
solar values, \iec [Fe/H] $\equiv \log (N_{\rm Fe}/N_{\rm H}) - \log
(N_{\rm Fe}/N_{\rm H})_{\solar}$.  We note that the
metallicity is a highly nonlinear measure of time.
}
If the Be was produced by the interaction of energetic
cosmic ray protons and alpha particles with ISM C and O and/or
by that of cosmic ray C and O {\it accelerated out of the ISM\/} on
ISM protons and alpha particles, the Be {\it rate} of
production would be proportional to the metallicity of the ISM, so
that its abundance would be expected to increase quadratically with
galactic metallicity.
However, if the bulk of all cosmic rays are accelerated from fresh
ejecta material and energetic C and O nuclei produce Be by interacting
with protons and helium in the ISM, the production of Be would be
approximately independent of global metallicity, and a linear increase
of its galactic abundance would be expected.
We note that the observational situation regarding the galactic
evolution of Be/H with O/H is far from settled since there are
indications that O/H does not increase as rapidly as Fe/H in the early
Galaxy (\egc Fields \& Olive 1999, based on data by Israelian,
Garc\'\i a L\'opez, \& Rebolo, 1998 and Boesgaard \etal 1999).  If
this is the case, the problem with Be is lessened since C and O are
the progenitors of Be, not Fe.

Of course not all models for cosmic ray origin assume that the
observed enhancement of refractory elements results from material
drawn from dust (see, for example, Shapiro 1999; Yanagita \&
Nomoto 1999).  We believe, however, that a consensus for dust is
emerging and do not consider alternative models here.  Instead, we
assume the refractory material comes from dust, and that efficient
diffusive shock acceleration in SNR shocks produces cosmic rays and
concentrate on the question of whether the dust originates in the
nearby interstellar and/or circumstellar material, or from grains
condensed within the fresh SN ejecta material.  We review some of the
physics associated with nonlinear diffusive shock acceleration and
with the acceleration of dust grains, in order to compare the ISM and
fresh ejecta models for cosmic ray origin (see Meyer \& Ellison
1999, in this volume, for a detailed study of the currently observed
cosmic ray composition).

\section{Specific Fresh Ejecta Models}

A fundamental problem exists for any model that wishes to produce the
{\it bulk} of the cosmic rays from fresh SN ejecta material since, for
any given SNR (even those in the early galaxy), the forward shock will
process far more ISM and/or circumstellar material than resides in the
fresh ejecta.
If we assume a simple Sedov solution for a supernova (SN) of explosive
energy
$\EnSN = 1\x{51}$ erg in a uniform ISM of hydrogen density $\nH \sim 1$
\pcc,
we estimate after a typical SN lifetime of
$\tSNR \sim 10^5$ yr that
$\Vsk \sim 120$ \kmps,
$\Rsk \sim 30$ pc,
and
\label{\mswept}
$$
\Mswept \sim
(4/3) \pi \Rsk^3 \, \rho
\sim 4\x{3} \,
\Msun
\ ,
\eqnum
$$
for the mass of material swept up by the forward shock (\egc Sedov 1959).
Here, $\Vsk$ is the shock speed, $\Rsk$ is the shock radius,
and a ratio He/H $= 0.1$ by number is assumed.  Compared to
this figure, the mass of the ejecta (typically $\sim 10\Msun$ for Type
II SN) is tiny.  We note that even if $\nH = 10^{-3}$ \pcc, a Sedov
solution after $10^5$ yr gives $\Mswept \sim 250 \Msun$.
Thus, any scenario which tries to generate most cosmic rays
from fresh ejecta must contain some mechanism for producing a large
enhancement, of order $30-500$, in the accelerated population of
ejecta material over swept up material.
This enhancement might come, either from preferential
injection and acceleration of ejecta material, and/or from a
particular trait of the SNR environment, which could be highly
enriched in ejecta material {\it before} the SN explodes.  Two recent
models attempt to do this, and we discuss them now.

\subsection{Isolated SNR Model}

The isolated SNR model of \LRK (1998) suggests that the bulk of
present day cosmic rays originate from the acceleration of fresh
ejecta material
located {\it within\/} the forward shock.
Problems we see with this model include:

\newlist
\list In order to avoid adiabatic losses and because the forward shock
contains far more energy than the reverse shock, this model requires
that dust grains formed in the SN ejecta reach the forward shock,
where they sputter, and that the sputtered material is then
accelerated by the forward SNR shock.
For this to occur, the newly formed grains must remain essentially
{\it uncoupled} to the local plasma, so they can move fast enough to
overtake the shock, which starts off with a much higher speed than the
grains, but later slows down as it sweeps up ISM material (\egc
Chevalier 1982).

We can estimate the degree of coupling which will allow grains to
reach the shock by assuming that grains of speed $\Vg$ diffuse without
slowing down with a mean free path $\Lgrain \sim f \, \Rsk$, where
$\Rsk$ is the shock radius and $f=1$ implies uncoupled, ballistic
motion.  We note that in order for diffusive shock acceleration to
occur,
the ratio of the diffusion length to the shock radius,
$\DiffL/\Rsk \sim \kappa/(\Vsk\Rsk)$ must be less than one,
and is typically assumed to be $<0.1$ for the highest energy particles
accelerated, and far less for lower rigidity particles.  Here, $\kappa
= (1/3) \Lgrain\, \Vg$ is the diffusion coefficient.  This implies
that $\Lgrain < 0.3 (\Vsk/\Vg) \Rsk$ or $f<0.03$ for grains to be
accelerated to a modest $\Vg = 10 \Vsk$.  If the grains execute a
random walk, they will diffuse a mean distance $\Rsk \sim \sqrt{N}
\Lgrain$ after $N$ scatterings or after a time, $\tG \sim N t_c = N
\Lgrain/ \Vg$, where the collision time $t_c = \Lgrain/\Vg$.
Therefore, the time for a grain to reach $\Rsk$ is
$
\tG \sim \Rsk / (f \Vg)$.
If we assume the forward shock moves as in a standard Sedov
solution in a uniform external medium,
$$
\Rsk \simeq \left ( { \EnSN \over \rho} \right )^{1/5} \,
\tSNR^{2/5}
\ ,
\eqnum
$$
where $\rho = 1.4
\mp \nH$ is the
ISM density. Equating these
times and eliminating $\Rsk$, we have an estimate for
the time it takes for the grain to overtake the shock in terms of the
coupling parameter $f$:
$$
\tG \sim
(f\, \Vg)^{-5/3} \,
\left ( { \EnSN \over \rho} \right )^{1/3}
\ ,
\eqnum
$$
or,
$$
\tG \sim 
285 \,
\left ( {f \over {0.03}}\right )^{-5/3}
\left ( { \Vg \over {c}}\right )^{-5/3}
\left ( {\nH \over {{\rm cm}^{-3}}}\right )^{-1/3}
\left ( {\EnSN \over {10^{51}{\rm erg}}}\right )^{1/3}
\ {\rm yr}
\ .
\eqnum
$$
This can be compared to the momentum loss time scale,
$\tloss$, from friction for a grain of size $a$ and mean molecular
weight $\mu$ as given in \EDM (1997),
\label{\tlossEq}
$$
\tloss \simeq
8 \,
\mu \,
\left ( {a \over {10^{-7}{\rm m}}}\right ) \,
\left ( {\nH \over {{\rm cm}^{-3}}}\right )^{-1} \,
\left ( {\Vg \over {c}}\right )^{-1} \,
\ {\rm yr}
\ .
\eqnum
$$
%
The ratio, $\tG/\tloss$ must be considerably less than 1 for grains
to easily reach the forward shock without slowing but,
$$
{\tG \over {\tloss}} \sim
30 \,
\left ( {f \over 0.03} \right )^{-5/3} \,
\left ( {\mu \over {56}}\right )^{-1} \,
\left ( {a\over {10^{-7} {\rm m} }} \right )^{-1}
\left ( {\Vg \over {10^3 \, {\rm km \, s}^{-1}}}\right )^{-2/3}
$$
$$
\left ( { \nH \over {{\rm cm}^{-3}}}\right )^{2/3}
\left ( {\EnSN \over {10^{51} {\rm erg}}}\right )^{1/3}
\ ,
\eqnum
$$
showing that for typical parameters,
$\tG/\tloss > 1$.
We conclude that it is unlikely that a significant fraction of
grains will reach the forward shock unless $f > 0.1$, \iec unless they
are uncoupled and do not interact significantly with the background.

It has been suggested that the \al26 $\gamma$-ray line observations of
Naya \etal (1996) indicate Doppler broadening from a speed of $\sim
450$ \kmps, and that this gives some support for fast grains
getting outside of SNR forward shocks.  However, the connection to
{\it fresh ejecta} grains is by no means conclusive.  First of all,
the balloon observations have poor statistics and need to be confirmed
(the line may not, in fact, be broader than that produced by galactic
rotation).  Next, the observations have a wide field of view (20 or 30
degrees or more) meaning that a large fraction of the Al containing
dust in the ISM must retain a high speed for $\sim 10^6$ yr after
ejection.  Using equation~(\tlossEq) to estimate the momentum-loss
timescale for Al grains ($\mu=26$) colliding with the background gas,
we find for $\nH=1$ \pcc, that a grain with $a=10^{-7}$ m will slow
from $500$ \kmps\ in $\sim 10^5$ yr, considerably shorter than the
$1.04\x{6}$ yr mean lifetime of \al26\ (\egc Fuller \etal 1982),
implying that reacceleration must take place.  Finally, there are
alternative explanations for fast grains other than that they retain
the speed of ejection, \egc shock acceleration in a single event, as
described in \EDM (1997), or reacceleration of ISM grains by multiple
shock encounters (Sturner \& Naya 1999).

There have been calculations of the Raleigh-Taylor instability in
young SNRs (\egc Jun \& Norman 1996) that suggest that some clumps of
fast-moving ejecta may punch through the outer
shock. However, there are good reasons to believe that these are
relatively minor effects (see Drury and Keane 1995).

\list The \LRK (1998) model requires that the grains sputter at the
forward shock, thus releasing ions that are then accelerated to cosmic
ray energies.  Sputtered ions are expected to retain the speed of the
grain so, if the ions encounter the shock, they will be injected and
accelerated more efficiently than thermal particles.
This type of enhanced acceleration is described in
\EDM (1997) for ISM grains which are picked up and accelerated by the
forward shock.
\LRK suggest the same mechanism can
provide the required enhancement of refractory materials seen in
cosmic rays using fresh ejecta grains only.
However, considering that the fresh ejecta
grains must be weakly coupled to the
plasma ($f \gtrsim 0.03$) in
order to reach the forward shock at all, it is not clear how they can
interact strongly and sputter at the shock to produce a significant
yield of refractory ions.
Now, it's possible to assume that sputtering may occur at the
forward shock, and not elsewhere, if the grains start to scatter
there efficiently because of the increase in magnetic turbulence and
density created by the shock compression.  But, both the
turbulence and the density increase occur predominantly {\it behind\/}
the shock.
Clearly a quantitative estimate is required, but qualitatively we
guess that the yield of sputtered ions accelerated to cosmic ray
energies is small, for two reasons:
{\it (i)\/} because the sputtering process only occurs
over the duration of the reverse shock, which is a small
fraction of the lifetime of the
forward shock,\footnote{The lifetime of the reverse shock should be
within a few factors of
$
t_c \simeq 0.36 [\Mej^5/(\EnSN^3\, \rho^2)]^{1/6}
\sim 200
$
yr
for standard parameters
(Chevalier 1982).
As explained in
Lingenfelter, Ramaty, \& Kozlovsky, new grains are formed only as long as
there
exists cold material, \iec before the reverse shock has heated the
entire ejecta material and collapsed at the center of the remnant.
Since the reverse shock lifetime, $t_{\rm RS}$, is a small fraction of
the forward shock lifetime, $t_{\rm FS}$, an estimate of the upper
limit on the fraction of the energy available to the forward shock
that can be expected to be available for acceleration of {\it
ejecta\/} grains is on the order $(t_{\rm RS}/t_{\rm FS}) \sim 0.01$
only.}
and {\it (ii)\/} because the ions that are sputtered from the
grains will be produced mostly behind the shock, where they have a
small probability of ever reaching the shock and being further
accelerated. If only those ions sputtered in front of the shock are
accelerated (as is most likely) and these represent only a small
fraction of the total sputtered ions, the
yield will be small.

\list Finally, no connection is made in this model between the
acceleration of
refractory (from fresh ejecta grains) and volatile cosmic ray
elements,
including, of course, H and He.  Since H and He and the other volatile
elements are not locked in grains, these elements are presumably
accelerated by an independent mechanism, \iec by the forward shock
sweeping up the ambient ISM. Since the ejecta grains only interact at
the forward shock for a brief time, while the forward shock will
presumably accelerate the ISM over its entire lifetime, any
quantitative estimate from this model would show, we expect, that
refractories are strongly {\it underabundant} relative to volatiles, in
disagreement with observations.
However, since
\LRK assume that the speed of the fresh ejecta grains, prior to
encountering the shock, is essential to
provide superthermal sputtered ions which will then be preferentially
injected and accelerated, the forward shock can
preferentially accelerate sputtered ions from the ejecta grains which
manage to overcome it, but will not do so for the ambient ISM grains
which are at rest. It is not clear what this model might predict for
the ratio of volatile to refractory cosmic rays and it is certainly
not obvious that quantitative estimates would match the observations.

\subsection{Superbubble Models in Today's Galaxy}

In view of the difficulties for an isolated SNR to accelerate its own
fresh ejecta material, several authors have considered the possibility
that SNRs within an OB association forming a superbubble may
accelerate the trapped local superbubble medium, which is enriched in
heavy elements by the ejecta of {\it previous\/} SNae, and by the
winds of the Red Giants and Wolf-Rayet stars of the OB association
(Bykov \& Fleishman 1992; Bykov 1999; Parizot \etal 1998; Parizot
1998; Higdon \etal 1998; Parizot \& Drury 1999).
Such models, if applied to the early Galaxy, may represent an
alternative way of explaining a ``primary''
origin for the galactic Be.
%
%
They may also circumvent our objections to a direct
acceleration, by a particular SNR, of its own, internal, SN ejecta
material: a forward SNR shock expanding within the superbubble,
indeed, will simultaneously accelerate the external, possibly
enriched, gas and dust in the superbubble.
However, despite these advantages, we believe there are still concerns
that require quantitative estimates, before superbubble models
wishing to explain the bulk of present day cosmic rays can be
accepted. These include:

\newlist
 \list
Since the material throughout the whole of the superbubble is not
dominated by fresh ejecta, but comes mainly from material evaporated
from the walls, \HLR (1998) give arguments explaining why the
majority of supernovae are likely to explode in a central `core',
which is dominated by fresh ejecta.
However, the core will only be dominated by refractory material if the
fast, newly formed grains are {\it coupled} to the plasma
magnetically, \iec $f \ll 1$.  That this is so can be seen directly:
grains with speeds $\Vg \sim 2000$ \kmps\ will move ballistically
about 600 pc during the mean time $t_{\rm SNOB} \sim 3\x{5}$ yr
between two SN explosions within the superbubble. The time between
SNae is assumed by \HLR (1998) to avoid problems of accelerating
electron-capture nuclei too soon after their nucleosynthesis (\egc
Lukasiak \etal 1997; Connell \& Simpson 1997; Binns 1999).  Since 600
pc is larger than the core size ($\le 400$ pc) estimated by Higdon,
Lingenfelter, \& Ramaty, the core will be {\it depleted} in refractory
material, unless the grains actually diffuse at a much slower rate.
While we believe it is likely that the grains are magnetically coupled
to the plasma and diffuse slowly, it must be noted that the \HLR model
requires that they are coupled, while the \LRK model requires that
they are not (except in the immediate vicinity of the shock).  Of
course, detailed calculations might show that grain scattering is more
intense in the strong turbulence present in superbubbles than in the
post-forward shock environment of an isolated  SNR.

 \list
Highly volatile elements in {\it current\/} cosmic rays show a mass
dependent enhancement which goes approximately as $\propto A^{0.8\pm
0.2}$ (\MDE 1997), and this has been shown to be a natural result of
the nonlinear shock acceleration of gas ions by \EDM (1997).
For a low temperature ISM at $T \sim 10^4$ K photoionized by SN and/or
stellar UV's, the charge states of all the volatiles are $Q\sim $1 or
2, so that $A/Q$ = (1 to 2)$\cdot A$, \ie roughly $A/Q \propto A$.
The observed composition thus requires that the acceleration
efficiency is approximately proportional to $(A/Q)^{0.8}$.
This result is consistent with acceleration predominantly by {\it
weak shocks\/} (Berezhko, Elshin, Ksenofontov 1996; \EDM 1997), which
seems indicated by the evolution of SNRs (\iec most ISM material is
processed when shocks are old and large), the cosmic ray spectrum, and
the cosmic ray hydrogen abundance (see \EDM 1997).
Superbubbles, however, are presumed to be hot ($T>10^6$ K) and diffuse
($n \le 10^{-3}$ \pcc).  If $T \sim 10^6$ K, the ion charge states are
such that $A/Q\sim A^{0.4}$,
%
%
so that the observed enhancements $\propto A^{0.8}$ require that the
acceleration efficiency is $\propto (A/Q)^{2.0}$.
This is possible only if {\it strong shocks\/} are predominant in
the acceleration, which seems to conflict with
the
requirement that shocks die out {\it
before\/} they expand beyond the ejecta enriched core of the
superbubble, which requires very low Mach number shocks, \iec shocks
in an extremely hot plasma.

A possible way around this problem involves recent ISO observations of
CAS A showing fast moving knots (Arendt \etal 1999). If volatiles are
present in these knots and evaporate off, they will have $Q \sim 1$ or
2, giving the needed $A/Q$ dependence in the hot medium. However,
Arendt \etalc estimate that the mass in the dust and knots is only
$\sim 3.8\x{-2} \Msun$, a small fraction of the ejecta mass.

If $T\gg 10^6$ K in the superbubble (\egc Mac Low \& McCray 1988), most
volatiles will have $A/Q\sim 2$,
%
%
and the observed increased enhancements of heavier volatiles
will not be produced.
A problem therefore results if the shocks are required to die out {\it 
before\/} they expand beyond the ejecta enriched core of the 
superbubble, which requires very low Mach number shocks, \iec shocks 
in an extremely hot plasma.

We conclude that a predominant origin of {\it current\/} cosmic rays
in superbubbles is inconsistent with the observed composition of the
cosmic ray volatile elements.
This, however, does not preclude a predominant origin of cosmic rays
in superbubbles {\it in the early Galaxy.\/} We, indeed, do not know
what the cosmic ray composition was at that time!

 \list
Another problem with the superbubble model of Higdon \etal (1998)
concerns the small mass of superbubble material processed by the
forward shock.  If we take the parameters assumed by Higdon \etal for
a superbubble SN explosion, \iec
$\nH \sim 10^{-3}$ \pcc,
$\Vsk \sim 2000$ \kmps,
$\Rsk \sim 40$ pc, and
$\tSNR \sim 10^4$ yr,
we find $\Mswept \sim 14 \Msun$.
Comparing this to the mass of $\sim 4000 \Msun$ swept up by a typical
isolated SNR in the ISM (eq.~\mswept) suggests that SNRs confined to
low density, enriched superbubble cores contribute little to the
cosmic ray mix.

We note that if the shocks are not required to remain small
(necessitated by wanting fresh ejecta material to dominate), extended
superbubble models may contribute substantially to the CR mix.  In the
model of Bykov \etal (Bykov \& Fleishman 1992; Bykov 1999), an
ensemble of many weak shocks and magnetic turbulence is formed in a
superbubble by the combined action of strong stellar winds and SN
explosions.  The accelerated particle spectrum that results is quite
different from that expected from only strong shocks, being flatter at
nonrelativistic energies (from second-order Fermi acceleration off
magnetic turbulence) and steeper at relativistic energies (dominated
by first-order acceleration but steep because of the low Mach number
of the SN shocks in the hot bubble at later stages).  With a suitable
choice of parameters (both for the superbubble and for galactic
propagation), this spectral shape can, however, be made
reasonably consistent with CR spectral observations.
Furthermore, since the acceleration is not restricted to the
ejecta-enriched core, the cosmic ray material can be a relatively
normal mixture of ISM gas and dust and, except for the $A/Q$
dependence just discussed, should be reasonably consistent with the
composition observations.  The great advantage of extended superbubble
models is that they may be able to supply a cosmic ray mix in the early
galaxy that is consistent with the metallicity dependence of Be (Bykov
1995).

\section{Acceleration of Cosmic Rays From the Interstellar Medium Mix}

We believe that a simple way to overcome all of the above problems
with producing present day cosmic rays is to assume that forward SNR
shocks accelerate external gas and dust. This, of course, will not
resolve the questions concerning the primary Be in the early
Galaxy. The details of this model are given in
\EDM (1997) and we summarize the basic ideas here.

\newlist
 \list
Forward SNR blast waves {\it simultaneously} accelerate external, \iec
ISM and/or circumstellar, gas and dust;

 \list
The volatile elements, as gas ions, are accelerated directly by the
shock.  Since nonlinear effects from efficient acceleration produce a
smooth shock structure, the volatiles obtain an enhancement,
consistent with observations, which is an increasing function of $A/Q$
or mass.  The high observed H/He ratio
can be accommodated in this model if the shock Mach number is low;

\list
The charged dust grains are assumed to be coupled to the plasma and to
act exactly as ions with the same rigidity.  Dust grains have large
$A/Q$ ratios on order $10^{8}$, hence large rigidities even at very
low speeds, but not larger than those of $\sim 10^{13-15}$ eV protons,
which are assumed to be accelerated by SNR shocks.  We assume that
dust grains pitch-angle scatter off the same magnetic field
irregularities as protons of the same rigidity (\iec gyroradius).  The
only difference is that dust grains of a given gyroradius will have a
much smaller speed than a proton with the same gyroradius. But as long
as the speed of the dust grain is greater than the speed of the
magnetic irregularities (\iec the \alf velocity), there should be no
fundamental difference with the wave-particle interaction.  The
essential point is that, if SNR shocks can accelerate protons to
$10^{13-15}$ eV, they can accelerate dust grains with typical
properties to $\gtrsim 0.1$ MeV/nucleon, with a high efficiency due to
the large grain $A/Q$.
At these energies, friction on the gas will actually prevent
further acceleration.
The gyroperiod, gyroradius, and speeds of typical grains are shown in
\EDM to be consistent with assumptions of the model;

\list
As the dust grains diffuse back and forth across the shock, they
sputter against the background plasma, and the ions sputtered off
outside the shock are later overtaken by the expanding shock and are
further accelerated.  Since these sputtered ions have a superthermal
speed, approximately that of the grain, they are accelerated much more
efficiently than thermal gas ions.  Combining the sputtering and
acceleration efficiencies allows a direct calculation of the absolute
acceleration efficiency of the refractory material and, since the
grain acceleration and sputtering processes continue throughout the
lifetime of the forward shock, the yield of refractory cosmic rays is
large;

\list
The same physical processes in the same shock apply
jointly to the gaseous and refractory materials, allowing a direct
determination of the relative acceleration efficiency of
%
%
volatile and refractory elements, accelerated as gas ions, and
from dust grains, respectively.  Refractories obtain a net enhancement
over volatiles, consistent with observations.
At variance to the volatiles, the enhancements for the refractories
are found largely mass independent: in the crucial, early acceleration
stages, they are, indeed, accelerated as constituents of entire
grains, {\it not\/} as individual ions. This near mass independence is
also indicated by the observations;

\list
The mechanism works for all types of supernova remnants, so it
naturally {applies for the most massive SNae, the explosions of which
have been preceded by the ejection of a major fraction of the stellar
mass (largely processed!)  in the form of winds during the Wolf-Rayet
phase.  The subsequent SNR shock acceleration of this pre-SN wind
material should account for the present day observed \Ne22\ and C
excesses in cosmic rays (for C, partial trapping in grains also adds
to the observed excess);

\list
The observed O excess by a factor of $\sim 3$ relative to neighboring
volatile species N and $^{20}$Ne, can be explained in a fully
consistent way by noting that 15--20\% of O is trapped in silicates,
the principal constituents of grain cores. This grain material is
accelerated $\sim 10$ times as efficiently as the gas-phase O,
yielding an overall enhancement of a factor $\sim 3$;

\list
Finally, only standard assumptions for grain properties (size, charge,
friction, sputtering, etc.)  and nonlinear diffusive shock
acceleration are required to produce a quantitative model that
adequately matches all of the galactic cosmic ray abundance data to
within experimental uncertainties.  While our results are still model
dependent and depend on the parameters we chose for grain size,
sputtering rates etc., they do not depend sensitively on any of
these parameters.  Furthermore, all of the underlying assumptions of
diffusive shock acceleration have been confirmed by direct tests
against spacecraft observations of shocks within the heliosphere (\egc
Lee 1982; \EMP 1990; Baring \etal 1997).

\section{Conclusions}

Based on the strong evidence from observations (reviewed by Meyer \&
Ellison, this issue) that fresh ejecta does not match the current
cosmic ray composition, and on the difficulties with fresh ejecta
acceleration models, compared with the success of an alternative model
that accelerates external (ISM and/or circumstellar) material, we
conclude that the linear increase in Be abundance with metallicity
observed in halo stars cannot imply that a significant fraction of the
cosmic rays {\it seen today} come from fresh ejecta.

This means that either, {\it (i)\/} cosmic ray production in
the early galaxy was quite different from today, so that fresh ejecta
did provide the bulk of the cosmic rays produced then,
or {\it (ii)\/} that cosmic rays in the early galaxy were
produced essentially as they are today, but that the {\it
comparatively few\/} cosmic rays accelerated out of freshly
synthesized material were dominant as regards Be production, since
virtually no Be could be produced out of any material (cosmic rays, or
at rest) originating in the then metal poor ISM.
It remains to be seen if {\it (ii)\/} can provide a quantitative
explanation of the Be abundance observations.

\acknowledgments

We thank the organizers for arranging a very interesting meeting.
This work was supported, in part, by NASA's Space physics Theory
Program. The authors thank A. Bykov, L. Drury, and E. Parizot for many
helpful comments.

\def\reff{\par\noindent \hangafter=1 \hangindent=0.8truecm}
\def\aa#1#2#3{ 19#1, {\it A.A.,} {\bf #2}, #3.}
\def\aasup#1#2#3{ 19#1, {\it A.A. Suppl.,} {\bf #2}, #3.}
\def\aj#1#2#3{ 19#1, {\it A.J.,} {\bf #2}, #3.}
\def\anngeophys#1#2#3{ 19#1, {\it Ann. Geophys.,} {\bf #2}, #3.}
\def\annrev#1#2#3{ 19#1, {\it Ann. Rev. Astr. Ap.,} {\bf #2}, #3.}
\def\apj#1#2#3{ 19#1, {\it Ap.J.,} {\bf #2}, #3.}
\def\apjlet#1#2#3{ 19#1, {\it Ap.J.(Letts),} {\bf  #2}, #3.}
\def\apjpress{{\it Ap. J.,} in press.}
\def\apjs#1#2#3{ 19#1, {\it Ap.J.Suppl.,} {\bf #2}, #3.}
\def\asr#1#2#3{: 19#1, {\it Adv. Space Res.,} {\bf #2}, #3.}
\def\araa#1#2#3{ 19#1, {\it Ann. Rev. Astr. Astrophys.,} {\bf #2},
#3.}
\def\ass#1#2#3{ 19#1, {\it Astr. Sp. Sci.,} {\bf #2}, #3.}
\def\eos#1#2#3{ 19#1, {\it EOS,} {\bf #2}, #3.}
\def\icrcplovdiv#1#2{ 1977, in {\it Proc. 15th ICRC(Plovdiv)},
{\bf #1}, #2.}
\def\icrcparis#1#2{ 1981, in {\it Proc. 17th ICRC(Paris)},
{\bf #1}, #2.}
\def\icrcbang#1#2{ 1983, in {\it Proc. 18th ICRC(Bangalore)},
{\bf #1}, #2.}
\def\icrclajolla#1#2{ 1985, in {\it Proc. 19th ICRC(La Jolla)},
{\bf #1}, #2.}
\def\icrcmoscow#1#2{ 1987, in {\it Proc. 20th ICRC(Moscow)},
{\bf #1}, #2.}
\def\icrcadel#1#2{ 1990, in {\it Proc. 21st ICRC(Adelaide)},
{\bf #1}, #2.}
\def\icrcdub#1#2{ 1991, in {\it Proc. 22nd ICRC(Dublin)},
{\bf #1}, #2.}
\def\icrccalgary#1#2{ 1993, in {\it Proc. 23rd ICRC(Calgary)},
{\bf #1}, #2.}
\def\icrcrome#1#2{ 1995, in {\it Proc. 24th ICRC(Rome)},
{\bf #1}, #2.}
\def\icrcromepress{ 1995, {\it Proc. 24th ICRC(Rome)}, in press.}
\def\grl#1#2#3{ 19#1, {\it G.R.L., } {\bf #2}, #3.}
\def\jcp#1#2#3{ 19#1, {\it J. Comput. Phys., } {\bf #2}, #3.}
\def\JETP#1#2#3{ 19#1, {\it JETP, } {\bf #2}, #3.}
\def\JETPlet#1#2#3{ 19#1, {\it JETP Lett., } {\bf #2}, #3.}
\def\jgr#1#2#3{ 19#1, {\it J.G.R., } {\bf #2}, #3.}
\def\jpG#1#2#3{ 19#1, {\it J.Phys.G: Nucl. Part. Phys., }
{\bf #2}, #3.}
\def\mnras#1#2#3{ 19#1, {\it M.N.R.A.S.,} {\bf #2}, #3.}
\def\nature#1#2#3{ 19#1, {\it Nature,} {\bf #2}, #3.}
\def\nucphysB#1#2#3{ 19#1, {\it Nucl. Phys. B (Proc. Suppl.,}
{\bf #2}, #3.}
\def\pss#1#2#3{ 19#1, {\it Planet. Sp. Sci.,} {\bf #2}, #3.}
\def\pf#1#2#3{ 19#1, {\it Phys. Fluids,} {\bf #2}, #3.}
\def\phyrepts#1#2#3{ 19#1, {\it Phys. Repts.,} {\bf #2}, #3.}
\def\pr#1#2#3{ 19#1, {\it Phys. Rev.,} {\bf #2}, #3.}
\def\prD#1#2#3{ 19#1, {\it Phys. Rev. D,} {\bf #2}, #3.}
\def\prl#1#2#3{ 19#1, {\it Phys. Rev. Letts,} {\bf #2}, #3.}
\def\pasp#1#2#3{ 19#1, {\it Pub. Astro. Soc. Pac.,} {\bf #2}, #3.}
\def\revgeospphy#1#2#3{ 19#1, {\it Rev. Geophys and Sp. Phys.,}
{\bf #2}, #3.}
\def\rmp#1#2#3{ 19#1, {\it Rev. Mod. Phys.,} {\bf #2}, #3.}
\def\sp#1#2#3{ 19#1, {\it Solar Phys.,} {\bf #2}, #3.}
\def\ssr#1#2#3{ 19#1, {\it Space Sci. Rev.,} {\bf #2}, #3.}
%

\end{document}